\documentclass[doublespacing]{elsart}

\journal{Chemical Physics}

\newlength{\spaltenbreite}
\usepackage{amsmath, amsfonts, amssymb}
\usepackage{graphicx}
\begin{document}

\begin{frontmatter}

\title{Kinetics of quasi-isoenergetic transition processes
in biological macromolecules}

\author{E.~G.~Petrov, V.~I.~Teslenko}
\address{Bogolyubov Institute for Theoretical Physics, National Academy of Sciences of Ukraine, Metrologichna Street., 14-b, UA-03680, Kiev, Ukraine}

\date{\today}
\begin{abstract}
A master equation describing the evolution of averaged molecular state
occupancies in molecular systems  where alternation of molecular energy levels is caused by discrete dichotomous and trichotomous stochastic fields, is derived. This study is focused on the kinetics of quasi-isoenergetic transition processes in the presence of moderately high frequency stochastic field. A novel physical mechanism  for temperature-independent transitions  in flexible molecular systems is proposed. This mechanism becomes effective when the conformation transitions between quasi-isoenergetic molecular states take place. At room temperatures, stochastic broadening of
molecular energy levels predominates the energy of low frequency
vibrations accompanying the transition. This leads to a cancellation
of the temperature dependence in the stochastically averaged rate constants. As examples, physical interpretations of the temperature-independent onset
of P2X$_3$ receptor desensitization in neuronal membranes, as well as
degradation of PER2 protein in embrionic fibroblasts, are provided.\

\end{abstract}


\end{frontmatter}

\section{Introduction}
Transient processes in physical, chemical, and biological systems reflect the evolution of state occupancies associated with  different types of electronic, nuclear, and spin  degrees of freedom. From the physics point of view, the development of the  microscopic models that are able to shed light on possible mechanisms of transient processes,  is an important task.
Biological systems are characterized by complex molecular structure with fast, intermediate and slow motions of separate molecular groups.
Most transient processes in biosystems can be studied in the framework of phenomenological approach that is based  on the  equations with certain set of kinetic parameters (rate constants).   In these cases, however,  any experimentally observable transition process  occurs  simultaneously with the more fast transitions. For instance, the characteristic times of conformation transitions in  biopolymers (nucleic acids and protein chains) are several orders greater than the characteristic times of the  vibration relaxation. Therefore, the  phenomenological parameters including the  occupancies of conformation states and the rates of transitions between them, appear  averaged over  the much faster motions.
Numerous  experimental studies show that in biosystems, the temperature dependence of the reaction rates  exhibits  the Arrhenius like behavior, i.e.  an exponential increase of the  rate with temperature. However, there are  cases when a temperature dependence smoothes away or becomes anomalous (opposite to "normal"). Several  attempts were made to explain these effects in the framework of phenomenological approach, without exploring the microscopic properties of molecular structures \cite{oli95,sca97,cha07,zon07,sab07}. The goal of this study is to provide a quantitative microscopic description of both the smoothed away (temperature-independent) and normal (temperature-dependent) cases for the  multi-type reaction pathways arising within the complexity of molecular structure.

In biology, there are  three known types of reactions where microscopic effects could be essential. The first type is associated with the reduction-oxidation processes caused by the photoinduced electron transfer in the photosynthetic units of plants and bacteria. The basic mechanism of these reactions involves  electron tunneling between rigid redox centers like metal-containing porphyrin rings and quinones. The tunneling mechanism is supported by the absence of temperature dependence for the transfer rates in wide temperature regions including the room temperatures \cite{cha80,war89,feh98}. The physical model describing an electron transfer between redox centers is well known, and it is associated with the existence of strong coupling with local vibrational modes. The respective analytic expressions have been derived by Jortner \cite{jor76,jor80} who has generalized the classic Marcus's theory \cite{mar64} of transfer processes for a quantum case. When  contact conformations between the centers can be stabilized, an electron recombination rate averaged over  the conformation states,  has anomalous temperature dependence demonstrating a certain rise with the temperature decrease \cite{cha86}.

The second type of reactions is associated with the biochemical catalysis  in the  specific protein or metal-protein centers. As a rule, such catalysis leads to a notable structure reorganization in the active centers as well as  breaking of the old and creation of the new chemical or hydrogen bonds between the respective structure units. Catalytic process consists of various chemical steps that  include not only an electron and a proton transfer, but a two-electron, an electron-proton, a hydride or a hydrogen transfer as well \cite{ham01,pet03}. Conventionally, all such reactions are described in the framework of enzyme kinetics \cite{fer06} where the rate constants are calculated using the concept of an  activated complex strongly connected with the corresponding reaction coordinate. The motion of the reactants along reaction coordinates is accompanied by overcoming the activation barriers, so that the rate constants experience an activation  temperature dependence similar to the  Arrhenius's one. However, in the framework of Marcus's theory, a barrierless behavior becomes also possible,  if only the value of the reorganization energy would coincide  with the  driving force of the reaction (due, for instance, to a pronounced nuclear displacement in rigid complexes), thus cancelling the exponential temperature dependence. Such situation is probably realized during the binding of the NO molecule to the protoheme. The binding  appears as a kinetically mono-exponential process with approximately temperature-independent rate constant in the region of 200-290 K \cite{ye07}.

The third type of reactions is associated with the conformation transformations. Some of these reactions do not exhibit the Arrhenius's temperature dependence even in the range of room temperatures. For instance, it has been demonstrated that the closing rate for both the single-stranded DNA and the RNA hairpin-loop fluctuations, as well as the rate for the cyclic  $\beta $ -hairpin peptide folding,  are weakly dependent on temperature in the interval from 10 to 60$^{\circ}$C showing zero or even slightly negative enthalpies \cite{wal01,kuz07}. Another example of the temperature-independence in the gating kinetics of membrane proteins has been recently provided for the desensitization onset of P2X$_3$ purinoreceptors \cite{khm08}. The kinetics of this process is characterized by the two rate constants,  each being temperature-independent in the range of 25-40$^{\circ}$C. A very recent example is the observation of a single-exponential kinetics of temperature-insensitive (in the range of 27-37$^{\circ}$C), period-determining process in the mammalian circadian clock \cite{iso09}. Note that, in these cases, the experiments cover a physiologically important temperature range.

While physical mechanisms explaining both the temperature-independent electron tunneling and the barrierless ligand binding are more or less clear for two noted types of reactions, and can be well recognized in the framework of  Marcus-Jortner approach, it is not the case for the third one. Indeed, the conformational transformations in flexible molecular structures should involve activation of  low-frequency vibrations accompanying the quasi-isoenergetic configuration transitions. This would  make the temperature dependence of the transition rates exponential. However, several experimental results have demonstrated a presence of the temperature-independent  reaction pathways. Therefore, to  describe the quasi-isoenergetic transient processes, the theoretical approach has to be modified.

In general, the current theoretical description of transient processes in flexible molecules as well as in molecular chain structures employs two approaches. One is  based on quantum mechanics  and uses the recent {\it ab initio} methods (mainly, a density functional theory \cite{fri05}) along with the {\it ab initio} molecular dynamics simulations \cite{dol02}. Another approach involves the development of various physical models that allow one to describe the amplitude and the kinetic characteristics of the transient processes in terms of definite physical parameters like vibration coupling, energy bias, fluctuation frequency, level broadening etc. \cite{dat95}. Most of these parameters already appear in molecular Hamiltonians, others are described with the rate constants of the molecular transitions (see e.g. \cite{pet06,pmh05} and refs. there in). The problem  is, however, that dealing with large molecular systems like proteins or nucleic acids in aqueous or specific biological environment using {\it ab initio} methods is very expensive computationally. For instance, it takes 4 days using a cluster of four Linux-based computers to calculate a nanosecond molecular dynamic trajectory for a small protein containing  $\sim$5000 atoms and surrounded by $\sim$15000 water molecules \cite{kar05}. An overwhelming majority of biomolecular transient processes  requires times ranging from milliseconds to seconds. So, correct computer simulation of these processes is currently inaccessible, even by  modern supercomputers, even in times from thousands to millions of years.

Meanwhile, the rigorous approach of statistical mechanics  allows to describe  the evolution of an open (interacting with the environment) molecular system in the formalism of nonequilibrium density matrix. This formalism has been successfully applied to describe exciton and charge transfer in different types of molecular systems \cite{val00,ep84,maku04}. The fundamental advantage of the density matrix formalism lies in  the unified description of transfer processes  by taking into account both the heat bath vibrations  and the stochastic energy fluctuations  \cite{pet96,goy97,gh05}. Furthermore, contribution from the stochastic effects can be analyzed exactly in a non-perturbation manner \cite{pet96,gh05,pt91,ptg94,pet98}. Another advantage of the noted formalism is that it allows one to reduce the description of  evolution behavior to the finding of the solution of a set of kinetic equations for state occupancies with precise rate constants. Therefore, the interpretation of the results in terms of the rate constants becomes clear and useful for both physicists and biologists.

In this paper,  we develop an approach  that is specially adapted  for description of transitions between quasi-isoenergetic molecular states, with use of nonequilibrium density matrix method. We argue that, for quasi-isoenergetic transitions, both low-frequency molecular vibrations and stochastic alternation of molecular energies by thermodynamic fluctuations, should be simultaneously  accounted for. We obtain rather uncomplicated expressions for transition rate constants in various limiting cases. We demonstrate that in these cases, different types of temperature behavior - from exponential temperature dependence of Arrhenius like type to the temperature independence, are  adequately reproduced. The examples  include  a physical explanation for the  temperature-independent onset of P2X$_3$ receptor desensitization in neuronal membranes \cite{khm08} and   temperature-insensitive, period-determining process in  mammalian circadian clock \cite{iso09}.

\section{Model and basic stochastic equation}

In molecular systems, any transition process contains contributions from  different types of vibrations. Some of them define the common electron-vibrational molecular states involved in the transitions while the others  can be referred as a heat bath. Just owing to coupling with the heat bath, the transitions between the molecular states appear in the form of kinetic process. This process reflects an evolution of the molecular system towards the equilibrium, where  state occupancies satisfy the Boltzmann's relations. As it is well known from the physics of transfer processes in condensed matter, the stochastic fields created by interior motions in a dynamic system, can play an important role in evolution of state occupancies. Direct introduction of these fields into the  Hamiltonian of dynamic system significantly simplifies the consideration of the complex motions in the system. This  allows one to use only a limited number of general characteristics like amplitudes and frequencies of the stochastic field (see, for instance, refs. \cite{gh05,pt91,ptg94,pet98,sha78}). The stochastic approach  clarifies the physics of transition processes on the semi-phenomenological level. It becomes  especially important in the case of biological macromolecules  where a number of different types of nuclear motions, including the stochastic motions, are observed. It has been experimentally shown \cite{fay01} that structural  movements of separate polar residues in proteins create high frequency vibrations,  so that specific electric field fluctuations are present in a large number of biological macromolecules. Direct observation of fluctuation movements in biological molecules has been done with use  single-molecule spectroscopy. Namely, the observation that the fluorescence of a single flavin adenine dinucleotide cofactor at room temperature turns on and off indicating the stochastic enzymatic turnovers of a flavin reductase \cite{lu98,yan03}. The energy fluctuations at ambient temperatures have been also observed in a photosynthetic peripheral light-harvesting molecular complex of type 2 \cite{val07,jan08}. Thus, there exists an experimental evidence  supporting a stochastical approach to describe transition processes in biological macromolecules. The simplest way for such description is to incorporate the random processes in a stochastic Hamiltonian and, therefore, in the stochastic Liouville equation.

\subsection{Stochastic molecular Hamiltonian}

When the exact spatial molecular structure is unknown or uncertain (often in the case of biological macromolecules), the  modeling of the transition process has to be based on general physical principles. While developing such model, two  different types of motions in a macromolecular system have to be considered. First,  vibrations in a heat bath. Due to phonon exchange with a heat bath,  the transitions between the molecular states appear as the relaxation process where the phonon energy covers the difference between the energies of these states. Second, high frequency vibrations. The energy of the corresponding phonons exceeds the noted energy difference signigficantly. This makes a phonon exchange ineffective. However, high frequency motions create interior random fields that alternate the molecular energies. The above examples show that at such alternation, the molecular energies can be stochastically time-dependent and, thus, the transitions between the molecular states (accompanied by creation and annihilation of low frequency phonons) occur on the background of randomly alternated energy levels. Taking this notion into account,  we consider a model where molecular energy levels exhibit stochastic alternations caused by interaction with the surrounding structural groups.  Rather,  the transitions between the energy levels are associated with the harmonic vibrations (phonons) of a heat bath.  Respective Hamiltonian of the entire system (molecule + bath) can be represented in the form
%
\begin{displaymath}
H_{entire}(t)=\sum_a\,\big\{E_a(t) +\sum_{\lambda}\,[
\kappa_{a\lambda}(\beta_{\lambda}^+ + \beta_{\lambda})+\hbar\omega_{\lambda}(\beta_{\lambda}^+\beta_{\lambda}+1/2)]
\big\}\, |a\rangle\langle a|
\end{displaymath}
\label{ham1}
\begin{equation}
+\sum_{a,b}\,V_{ab}\,(1-\delta_{a,b})\,|a\rangle\langle b|
\label{ham1} \end{equation}
where $E_a(t)=E^{(0)}_a+\Delta E_a(t)$ is the energy of the $a$th molecular state ($E^{(0)}_a$ is the energy in the absence of an external time-dependent field, $\Delta E_a(t)$  is the energy variation caused by the stochastic fields), $\kappa_{a\lambda}$ is the parameter that characterizes the deviation of nuclei along $\lambda$th normal coordinate from the nuclear equilibrium position,  and $\omega_{\lambda}$ is the $\lambda$th vibration mode (phonon) belonging to the  bath Hamiltonian. Quantities  $\beta_{\lambda}^+$ and $\beta_{\lambda}$ refer to the operators of creation and annihilation of a phonon in a heat bath.  Thus, the entire system under consideration consists of two main parts. The first one is the dynamic quantum system (S) with the set of molecular states $\{|a\rangle\}$ and respective energies $\{E_a(t)\}$. The  second part represents a heat bath (B) with set of  vibrational modes $\{\omega_{\lambda}\}$. The transitions between the molecular states are determined by the term containing the matrix elements $V_{ba}$. In what follows, we represent the main parts of the Hamiltonian (\ref{ham1}) in completely diagonal form. To this end, it is convenient to use the Holstein's polaron-like transformation \cite{hol59,wag86} while defining the  respective unitary matrix in form
$U=\exp{\big(\sum_a\,\sigma_a|a\rangle\langle a|\big)}$ with $\sigma_a=\sum_{\lambda}\,g_a^{(\lambda)}\,(\beta_{\lambda}^+ -\beta_{\lambda})$ being the operator of the nuclear displacements in the $a$th molecular state ($g_a^{(\lambda)}\equiv \kappa_{a\lambda}/\hbar\omega_{\lambda}$ is the dimensionless coupling). Multiplying
the Eq. (\ref{ham1}) from the left by $U$ and from the right by $U^+$ we transform the starting Hamiltonian (\ref{ham1}) to
%
\begin{equation}
H(t)= H_S(t)+H_B +V\label{ham2}\,.
\end{equation}
Here,
%
\begin{equation}
H_S(t)=\sum_a\,(E_a(t)+\Delta E_a^{(pol)})\,|a\rangle\langle a|
\label{sham}
\end{equation}
is the molecular Hamiltonian where a proper energy of the $a$th state includes polaron shift     $\Delta E_a^{(pol)}=-\sum_{\lambda}\,\kappa_{a\lambda}^2/\hbar\omega_{\lambda}$. Bath Hamiltonian coincides with Hamiltonian of harmonic vibrations (phonons),
%
\begin{equation}
H_B=\sum_{\lambda}\,\hbar\omega_{\lambda}(\beta_{\lambda}^+\beta_{\lambda}+1/2)\,.
\label{bham}
\end{equation}
Off-diagonal part of entire Hamiltonian  appears now in the form
%
\begin{equation}
V=\sum_{a,b}\, (1-\delta_{a,b})\,\hat{V}_{ab}\,|a\rangle\langle b|
\label{trop}
\end{equation}
where $\hat{V}_{ab}= V_{ab}\,e^{\sigma_{ab}}$ and $\sigma_{ab}=\sigma_a-\sigma_b=\sum_{\lambda}\,g_{ab}^{(\lambda)}\,
(\beta_{\lambda}^+ - \beta_{\lambda})$. Operators $\hat{V}_{ab}$  characterize the transitions between different molecular states when the transitions are accompanied by the nuclear displacements (respective couplings  are $g_{ab}^{(\lambda)}=g_{a}^{(\lambda)}-g_{b}^{(\lambda)}$). Thus, the transitions between the molecular states are performed with the "phonon-dressed" off-diagonal interactions.

\subsection{Stochastic equation for a diagonal part of density matrix}

To perform the analysis of the experimental data,  a master equation for description of evolution of  the observable state occupancies $P_a(t)$, is reuired. Since the molecular energy levels $E_a(t)$ are the stochastic quantities,  this master equation appears as the equation averaged over random realizations of the energy shifts $\Delta E_a(t)$. We denote the averaging via the symbol  $\langle\langle...
\rangle\rangle$ so that $P_a(t)=\langle\langle{\mathcal
P}_a(t)\rangle\rangle$. Here, ${\mathcal P}_a(t)$ is the nonaveraged state occupancy which is defined as ${\mathcal P}_a(t)=\langle a|\rho (t)|a\rangle$ with $\rho (t)=tr_B\,\rho_{S+B}(t)$ being the molecular density matrix. This matrix is determined as a trace over the bath states on density matrix  $\rho_{S+B}(t)$ of the entire  quantum system  including the heat bath. The density matrix method assumes that irreversible (relaxation) transitions between the  quantum states of an open dynamic quantum system S are associated with an energy exchange between the system and the heat bath (via creation and annihilation of phonons). Since the coupling between the  molecular states and the heat bath is concentrated in off-diagonal interaction (\ref{trop}), it becomes important to derive a  master equation for  molecular occupancies in the form that allows one to control an expansion over the noted off-diagonal interaction. Due to the exact relation $\langle a|\rho (t)|a\rangle=\langle a|\rho_d (t)|a\rangle$,  an equation for diagonal part of molecular density matrix, $\rho_d(t)$, is quite sufficient. The details of the derivation  procedure  can be found, for instance, in refs.\cite{pet06,pmh05,lin91}. Following this procedure while taking into account  that, in the considered case, the molecular energies are stochastically time-dependent quantities, the  master equation for the $\rho_d(t)$ should appear in the stochastic form.

The starting point  is the stochastic Liouville equation
%
\begin{equation}
\dot{\rho}_{S+B}(t)=-\mathrm{i}\mathcal{L}(t)\rho_{S+B}(t)\label{leq1}\,
\end{equation}
where $\mathcal{L}(t)=(1/\hbar)\,[H(t),...]$ is the Liouville
superoperator associated with the Hamiltonian (\ref{ham2}). In line with
the  structure of this Hamiltonian, the $\mathcal{L}(t)$ consists of
two Liouville superoperators,
$\mathcal{L}_0(t)=(1/\hbar)\,[H_0(t),...]$ with $H_0(t)=H_S(t) +H_B$ and
$\mathcal{L}_V=(1/\hbar)\,[V,...]$ so that
$\mathcal{L}(t)=\mathcal{L}_0(t)+\mathcal{L}_V$.
Let introduce the
projection superoperators $\hat{T}_d$ and $\hat{T}_{nd}=I-\hat{T}_d$.
These will  expand any operator into its diagonal component (the matrix
elements are the occupancies) and its off-diagonal component (the matrix
elements are the coherences). The action of the operators $\hat{T}_d$ and
$\hat{T}_{nd}$ on the Liouville
equation (\ref{leq1}) generates a coupled set of differential
equations for diagonal and off-diagonal parts of the density matrix,
$\rho^{(d)}_{S+B}(t)=\hat{T}_{d}\rho_{S+B}(t)$ and $\rho^{(nd)}_{S+B}(t)=\hat{T}_{nd}\rho_{S+B}(t)$, respectively. These equations read
%
\begin{displaymath}
\dot{\rho}^{(d)}_{S+B}(t)=-\mathrm{i}\hat{T}_d\,
\mathcal{L}_V\rho^{(nd)}_{S+B}(t)\,,
\end{displaymath}
\begin{equation}
\dot{\rho}^{(nd)}_{S+B}(t)=-\mathrm{i}\hat{T}_{nd}\,
\mathcal{L}(t)\rho^{(nd)}_{S+B}(t)-
\mathrm{i}\mathcal{L}_V\rho^{(d)}_{S+B}(t)\,. \label{leq2}
\end{equation}
After substitution of the second equation into the first one, the following integro-differential equation for the  diagonal part of the density matrix is achieved
%
\begin{equation}
\dot{\rho}^{(d)}_{S+B}(t)=-\int_{0}^{t}dt'\,[\hat{T}_d\,
\mathcal{L}_VS(t,t')
\mathcal{L}_V\rho^{(d)}_{S+B}(t')]\, \label{deq}
\end{equation}
where
%
\begin{equation}
S(t,t')=\mathrm{e}^{-\mathrm{i}\int_{t'}^{t}d\tau\hat{T}_{nd}\,
\mathcal{L}(\tau)} \label{mk}
\end{equation}
is the stochastic evolution superoperator. Bearing in mind the definition of the molecular density matrix, one arrives to the equation
%
\begin{equation}
\dot{\rho}_d(t)=-\int_{0}^{t}dt'\,tr_B[\hat{T}_d\,
\mathcal{L}_VS(t,t')
\mathcal{L}_V\rho^{(d)}_{S+B}(t')]\,. \label{deq}
\end{equation}
In many practical cases, it is common that a characteristic time of transition processes, $\tau_{tr}$, significantly exceeds the characteristic time  of the vibration relaxation, $\tau_{vib}$. This  makes it possible to refer the vibrations to the  heat bath. Moreover, the validity of the inequality $\tau_{vib}\ll\tau_{tr}$ allows one to use a  factorization $\rho^{(d)}_{S+B}(t)=\rho_d(t)\rho_B$ where $\rho_B=
\exp{(-H_B/k_BT)}/tr_B\,\exp{(-H_B/k_BT)}$ is the bath equilibrium density matrix ($k_B$ and $T$ are the Boltzmann's constant and the absolute temperature, respectively). Thus, if the transition processes occur on the time scale of $\Delta t\gg\tau_{vib}$, one can describe these processes with the coarse-grained master equation
%
\begin{equation}
\dot{\rho}_d(t)=-\int_{0}^{t}dt'\,\hat{M}(t,t')\,\rho_d(t') \label{deq1}
\end{equation}
where
%
\begin{equation}
\hat{M}(t,t')=
tr_B[\hat{T}_d\,
\mathcal{L}_VS(t,t')
\mathcal{L}_V\rho_B]\, \label{relsup}
\end{equation}
is the stochastic transition superoperator.

\section{Master equation for the averaged molecular occupancies}

Eq. (\ref{deq1}) can be treated as the basic coarse-grained stochastic equation for the rigorous description of the transitions between the
molecular states on the time scale $\Delta t\gg \tau_{vib}$. Its form is quite suitable for an expansion procedure. Thus, if a description is restricted by Born approximation over the $V$ one has to set $V=0$ in the evolution superoperator (\ref{mk}). It yields
%
\begin{equation}
\dot{\rho}_d(t)=-(1/\hbar^2)\,\int_0^t dt'\, tr_B\big\{\hat{T}_d[V,U(t,t')[V,
\rho_d(t')\rho_B]U^+(t,t')]\big\} \label{me1}
\end{equation}
where we have introduced the zeroth order evolution operator $U(t,t')=\hat{T}\exp{[-(i/\hbar)\int^t_{t'}\,
d\tau (H_S(\tau)+H_B)]}$ ($\hat{T}$ is the Dayson's chronological operator).

\subsection{Stochastic kinetic equation for the state occupancies}

In what follows, we concentrate on the transition processes caused by a weak coupling of the molecular states to the bath vibrations (phonons) so that the "dressed" transition operators $\hat{V}_{ab}$ in off-diagonal interaction (\ref{trop}) are considered as perturbation (see also refs. \cite{pet06,gh05,gpm95,pmh04}). Using the definition
 $\mathcal{P}(a;t)
=\langle a|\rho_d(t)|a\rangle$, one derives the following stochastic equation for the occupancies
%
\begin{equation}
\dot{\mathcal{P}}(a;t)=-\sum_b\,\int_{0}^t dt'\,
\big[\mathcal{G}_{ab}(t,t')\,
\mathcal{P}(a;t') - \mathcal{G}_{ba}(t,t')\,
\mathcal{P}(b;t')
\big]
\label{me2}
\end{equation}
where kernel
%
\begin{equation}
\mathcal{G}_{ab}(t,t')=\frac{2}{\hbar^2}\,|V_{ba}|^2\,{\rm Re}\,\big[
e^{i\Omega_{ab}(t-t')}\,\mathcal{R}_{ab}(t-t')\,\mathcal{F}_{ab}(t,t')\big]\,
\label{kern}
\end{equation}
exhibits a stochastic behavior through the stochastic
functional ${\mathcal F}_{ab}(t,t')=\exp{\{(i/\hbar)\int_{t'}^t\,
d\tau[\Delta E_a(\tau)-\Delta E_b(\tau)]\}}$.  In Eq. (\ref{kern}),
$\Omega_{ab}=(1/\hbar)[(E^{(0)}_a +E_{a}^{(pol)})-(E^{(0)}_b +E_{b}^{(pol)})]$ is the transition frequency. Coupling to the heat bath is reflected in the correlation function $\mathcal{R}_{ab}(\tau)=tr_B\,\big(\rho_B\,\exp{[\sigma_{ab}(0)]}
\exp{[\sigma_{ab}(\tau)]}\big)$. If the bath is associated with the harmonic oscillations, then $\sigma_{ab}(\tau)=\sum_{\lambda}\,g^{(\lambda)}_{ab}\,
[b_{\lambda}^+e^{-i\omega_{\lambda}\tau}
-b_{\lambda}e^{i\omega_{\lambda}\tau}]$ and, thus
%
\begin{equation}
\mathcal{R}_{ab}(\tau)=e^{-G_{ab}(\tau)}
 \label{corb1}
\end{equation}
where quantity
%
\begin{equation}
G_{ab}(\tau)=\frac{1}{2\pi}\int_{0}^{\infty}\,
d\omega\frac{J_{ab}(\omega)}{\hbar^2\omega^2}\,
\big[\coth{\frac{\hbar\omega}{k_BT}}\,(1-\cos{\omega\tau})+{\rm i}\sin{\omega\tau}\,\big]
 \label{g1}
\end{equation}
is of basic importance for any calculations done on the spin-boson model \cite{maku04,gh05,pmh04,cal83,leg87,gri98,wei98}. In Eq. (\ref{g1}),
%
\begin{equation}
J_{ab}(\omega)=2\pi\,\sum_{\lambda}\,
(g^{(\lambda)}_{ab}\hbar\omega_{\lambda})^2\,\delta (\omega-\omega_{\lambda})
\label{sf}
\end{equation}
is the spectral function that includes an information on both the  vibrational structure of macromolecule and the  characteristics of  coupling to the bath vibrational modes.

Form (\ref{g1}) is quite suitable for estimation of rate constants in the case of a strong coupling to the bath. This work, however, is focused on the transition processes where a weak coupling  is realized between the molecular states and the phonons. Therefore, it is more convenient to submit the  correlation function in the form
%
\begin{equation}
\mathcal{R}_{ab}(\tau)=e^{-D_{ab}}\,Q_{ab}(\tau)
 \label{corb2}
\end{equation}
where quantity $D_{ab}=\sum_{\lambda}\,(g^{(\lambda)}_{ab})^2\,(2n(\omega_{\lambda})+1)$
specifies the Debye-Waller factor while function
\begin{equation}
Q_{ab}(\tau)=\int_{-\infty}^{\infty}\,d\omega\,
e^{i\omega\tau}\prod_{\lambda}\sum_{p_{\lambda}=-\infty}^{\infty}\,
I_{|p_{\lambda}|}(z_{\lambda})
\Big(\frac{n(\omega_{\lambda})}{n(\omega_{\lambda})+1}\Big)^{p_{\lambda}/2}\,
\delta [\omega-\sum_{\lambda}\,p_{\lambda}\omega_{\lambda}]
\label{g2}
\end{equation}
fixes the  time dependence of the correlation function. In (\ref{g2}), $I_{p}(z)$ is the modified Bessel function,  and  $n(\omega)= [\exp{(\hbar\omega/k_BT)}-1]^{-1}$ is the Bose distribution function.  Coupling to the phonons is concentrated in the parameter $z_{\lambda}\equiv 2(g^{(\lambda)}_{ab})^2
\sqrt{n(\omega_{\lambda})[n(\omega_{\lambda})+1]}$,  while index $p_{\lambda}$ indicates the number of phonons of the $\lambda$th mode that accompany the transition. The minimal number of phonons is equal to 1,  so that $\sum_{\lambda}\,p_{\lambda}\geq
\ 1$.

Supposing the small nuclear displacement along the $\lambda$th normal coordinate,  we consider a behavior of the correlation function at  $z_{\lambda}\ll 1$. This allows one to employ the asymptotic form $I_{|p_{\lambda}|}(z_{\lambda})\approx (z_{\lambda}/2)^{|p_{\lambda}|}/|p_{\lambda}|!$, which indicates that the main contribution to transitions comes from the single-phonon processes. Therefore, by setting $I_0(z\ll 1)\approx 1$ and $I_1(z\ll 1)\approx (z/2)$, one can see that the form (\ref{g2}) reduces to
%
\begin{equation}
Q_{ab}(\tau)=\sum_{\lambda}\,(g^{(\lambda)}_{ab})^2\,R_{\lambda}(\tau)\,,
\label{g3}
\end{equation}
where $R_{\lambda}(\tau)=n(\omega_{\lambda})\exp{(i\omega_{\lambda}\tau)}+
[n(\omega_{\lambda})+1]\exp{(-i\omega_{\lambda}\tau)}$ is the  single-phonon factor. Bearing in mind that, at weak nuclear  displacements, the role of the Debye-Waller factor is not important ($\exp{(-D_{ab})}\approx 1$), one can further  specify the stochastic equation (\ref{me2}) setting $\mathcal{R}_{ab}(t-t')\approx Q_{ab}(t-t')$.

\subsection{Kinetic equation for the averaged occupancies}

To derive the equations for the observable occupancies $P_a(t)$, one has to
average the stochastic equation (\ref{me2}). Here, we consider only the
case of fast variations for each random energy difference $\Delta
E_a(\tau) - \Delta
E_b(\tau)$, when the characteristic times of the
stochastic and the transition processes ($\tau_{stoch}$ and $\tau_{tr}$,
respectively) obey the inequality $\tau_{stoch}\ll\tau_{tr}$. This
inequality allows us  to use a decoupling procedure $\langle\langle
{\mathcal F}_{ab}(t,t'){\mathcal P}_a(t')\rangle\rangle=
\langle\langle {\mathcal F}_{ab}(t,t')\rangle\rangle
P_{a}(t')$. Moreover, a non-Markovian character of the transition process becomes not important. Actually, let us set $t=t'+\tau$ and rewrite $P_{a}(t')$ as $P_{a}(t-\tau)=\exp{(-\tau\frac{d}{dt})}P_{a}(t)$. On the time scale $\Delta t\sim\tau_{stoch}$ the factor $\exp{(-\tau\frac{d}{dt})}$ can be estimated as $\exp{(-\tau_{stoch}/\tau_{tr})}\approx 1$ and thus  $P_{a}(t')\approx P_{a}(t)$. This is also supported by the conclusion that in the Born approximation over the $V$ \cite{foota}, the delay processes do not introduce any  noticeable contribution to the transition kinetics  \cite{akh81}.

The estimation of the averages like
$\langle\langle {\mathcal F}_{ab}(t,t-\tau)\rangle\rangle
=F_{ab}(\tau)$ is given in numerous papers (see, e.g. refs.
\cite{goy97,gh05,pt91,sha78,bri74}). The results indicate that
independently of a specific form of quantities $F_{ab}(\tau)$, all
of them decay with their own characteristic
times,  $\tau_{stoch}$. Taking this fact into account and using the
condition $\tau_{stoch}\ll\tau_{tr}$, one can shift the upper
limit in the integral of non-Marcovian stochastic equation (\ref{me2}) from $t$ up to infinity. Thus, providing the averaging
over the fast random realizations, we  transform this equation into a balance like kinetic equation,
%
\begin{equation}
\dot{P}_a(t)=-\sum_b\,\big[K_{ab}P_a(t)-K_{ba}P_b(t)\big]
\label{me4}
\end{equation}
with the averaged transition rate constants
%
\begin{equation}
K_{ab}=\frac{2|V_{ab}|^2}{\hbar^2}\,{\rm
Re}\,\sum_{\lambda}\,|g^{\lambda}_{ab}|^2\,
\int_{0}^{\infty}\, d\tau
e^{i\Omega_{ab}\tau}\,R_{\lambda}(\tau)F_{ab}(\tau)\,.\label{rate1}
\end{equation}

To derive an analytic form for the rate constant, one has to
calculate  function $F_{ab}(\tau)$. To this end, let us consider the
Kubo-Anderson random process \cite{kub54,and54}. During such
process,  the jumps from one value to other happen at random time
and independently from one another with the average frequency $\nu$.
In our case, a stochastic alternation is associated with energy
differences $\Delta E_a(\tau)-\Delta E_b(\tau)=\hbar
\alpha_{ab}(\tau)$. We limit  ourselves by the case when
fluctuations of the $\alpha_{ab}(\tau)$ are independent of the
precise molecular states $a$ and $b$. This situation is adequate for
transitions in two-level systems and in flexible molecular chains
containing identical structure groups. If $\alpha_{ab}(\tau)=\alpha
(\tau)$, then $F_{ab}(\tau) =F(\tau)=\langle\langle
X(\tau)\rangle\rangle$ where
$X(\tau)=\exp{[i\int_{0}^{\tau}\,d\tau'\alpha (\tau')]}$ is the
stochastic functional. Below, two examples of Kubo-Anderson process
are given.

\underline{\textit{Dichotomous process (DP)}}.  During the DP a random quantity fluctuates between two equiprobable values $+\sigma_D$ and $-\sigma_D$ with a mean frequency $\nu_D$ so that $\alpha_{ab}(t)=\alpha_{DP}(t)=\pm\sigma_D$.
Bearing in mind the exact properties of the DP, so as
$\alpha_{DP}^2(t)=\sigma_D^2$, $\langle\langle
\alpha_{DP}(t)\rangle\rangle=0$, and $\langle\langle
\alpha_{DP}(t)\alpha_{DP}(0)\rangle\rangle=\sigma_D^2\exp{(-\nu_D|t|)}$,
and using the Shapiro-Loginov theorem \cite{sha78}
in the form of relation
$d\langle\langle\alpha_{DP}(\tau)X(\tau)\rangle\rangle/ d\tau =i\nu_D
d\langle\langle X(\tau)\rangle\rangle /d\tau
+i\sigma_D^2\langle\langle X(\tau)\rangle\rangle$, one obtains the following {\it exact} equation  (see also refs. \cite{goy97,pt91,ptg94,pet98})
%
\begin{equation}
\ddot{F}(\tau)+\nu_D \dot{F}(\tau)+\sigma_D^{2} F(\tau)=0\,.
\label{treq1}
\end{equation}
This equation is similar to the one for a harmonic oscillator with frequency $\sigma_D$ and damping parameter $\nu_D$. Respective  solution reads
%
\begin{equation}
F(\tau)=(k_1e^{-k_2\tau}-k_2e^{-k_1\tau})/(k_1-k_2)\,.
\label{dp1}
\end{equation}
The quantities
%
\begin{equation}
k_{1,2}=\nu_D/2\pm\sqrt{(\nu_D/2)^2-\sigma_D^2}
\label{dpr1}
\end{equation}
determine the above noted characteristic stochastic
times $\tau_{stoch}^{(1,2)}=({\rm Re}\,k_{1,2})^{-1}$. At high fluctuation frequency, owing to condition
$\nu_D^2\gg 4\sigma_D^2$ the rate $k_1$ strongly exceeds  the rate $k_2$. Therefore, at the time scale $\Delta t \gg \tau_{stoch}^{(1)}$, the characteristic function appears as $F(\tau)\approx \exp{(-\gamma_D\tau})$,
where
%
\begin{equation}
\gamma_D=\sigma_D^2/\nu_D
\label{dem1}\underline{}
\end{equation}
exhibits itself as the only damping parameter specifying the characteristic stochastic time $\tau_{stoch}=\gamma_D^{-1}$.
An approximation of random energy variations by the DP is well justified for a two-level system.

\underline{\textit{Trichotomous process (TP).}} The random energy
difference $\Delta E_{a}(t) - \Delta E_{b}(t)=\hbar \alpha_{ab}(t)$
exhibits itself in the peculiar form of  Kangaroo process
\cite{sha78,bri74} where the $\alpha_{ab}(t)$ fluctuates between the
three equiprobable values,
$\alpha_{ab}(t)=\alpha_{TP}(t)=0,\pm\sigma_T$ and $\langle \langle
\alpha _{TP} (t)\rangle \rangle =0$. At the TP, the fluctuation
frequency  $\nu_T$  is assumed to be independent of the values of
the  amplitudes and, thus, the TP represents a member of the
Kubo-Anderson processes. In this case, the Shapiro-Loginov theorem
for the rules of differentiation of the $k$th order correlations
reads $d\langle \langle \alpha _{TP}^{k} (t)X(t)\rangle \rangle/dt
=-\nu_T \langle \langle \alpha _{TP}^{k} (t)X(t)\rangle \rangle
+\nu_T \langle \langle \alpha _{TP}^{k} (t)\rangle \rangle \langle
\langle X(t)\rangle \rangle +\langle \langle \alpha _{TP}^{k}
(t)\dot{X}(t)\rangle \rangle $. Using this theorem and bearing in
mind the exact properties of the TP so as  $\langle \langle \alpha
_{TP} (t)\alpha _{TP} (0 )\rangle \rangle =(2\sigma_T^{2} /3)\exp
(-\nu_T |t|)$  and  $\alpha _{TP}^{3} (t)=\sigma_T ^{2} \alpha _{TP}
(t)$, for the characteristic function  $F(\tau)$  one obtains the
following {\it exact} equation
%
\begin{equation}
\dddot{F}(\tau)+2\nu_T \ddot{F}(\tau)+(\nu_T ^{2} +\sigma_T ^{2} )\dot{F}(\tau)+(2\nu_T \sigma_T^{2} /3)F(\tau)=0\,.
\label{treq}
\end{equation}
This equation indicates the presence of three characteristic stochastic times. But,  if the condition $(\nu_T ^{2} +\sigma_T ^{2} )^{3/2}\gg \nu_T\sigma_T^2$ is valid, then
on the time scale $\Delta t\gg (\nu_T ^{2} +\sigma_T ^{2} )^{-1/2}$ the characteristic function manifests itself as a single exponential drop so that  $F(\tau)\approx \exp{(-\gamma_T\tau})$. The decay parameter
%
\begin{equation}
\gamma_T=\tilde{\sigma}_T^2/\tilde{\nu}_T
\label{dem2}
\end{equation}
determines the characteristic stochastic time $\tau_{stoch}=\gamma_T^{-1}$
which now depends on the renormalized stochastic frequency $\tilde{\nu}_T=\nu_T/2 +\sigma_T^2/2\nu_T$ and the renormalized stochastic amplitude $\tilde{\sigma}_T=\sigma_T/\sqrt{3}$.

\section{Temperature-independence of rate constants}

Above examples show that the characteristic function $F(\tau)$ exhibits an exponential drop with the  stochastic characteristic times $\tau_{stoch}^{(j)}$, $(j=1,2,...)$. The number and the form of the  characteristic times are dictated by a precise stochastic process. [At a large difference between the amplitude and the frequency of the discrete process (DP or TP), the main drop is governed by the only decay parameters (\ref{dem1}) or (\ref{dem2}).] Substituting the characteristic function  $F(\tau)$  in Eq. (\ref{rate1}) and calculating the respective integral, one can achieve an analytic  expression for the averaged transition rate constant.

In the case of DP, the use of the exact expressions (\ref{dp1})-(\ref{dem1}) reduces the averaged rate constant to  (note $\gamma\equiv \gamma_D$, $\nu\equiv \nu_D$)
%
\begin{displaymath}
K_{ab}=\frac{2\pi}{\hbar^2}\,
\sum_{\lambda}\, |\chi_{ab}^{(\lambda)}|^2
 \{[n(\omega_{\lambda})+1]\,L(\gamma,\nu;\Omega_{ab} -\omega_{\lambda})
\end{displaymath}
\begin{equation}
+n(\omega_{\lambda})\,L(\gamma,\nu;\Omega_{ab} +\omega_{\lambda})
\}\,.
\label{rate2st}
\end{equation}
Here, the quantity $\chi_{ab}^{(\lambda)}=V_{ab}\,g_{ab}^{(\lambda)}$ appears  as the phonon assisting inter-state coupling. In what follows, we rewrite the expression  (\ref{rate2st}) in the form
%
\begin{displaymath}
K_{ab}=\frac{1}{\hbar^2}\,|V_{ab}|^2\,
\int_{0}^{\infty}d\omega \frac{J_{ab}(\omega)}{\hbar^2\omega^2} \{[n(\omega)+1]\,L(\gamma,\nu;\Omega_{ab} -\omega)
\end{displaymath}
\begin{equation}
+n(\omega)\,L(\gamma,\nu;\Omega_{ab} +\omega)\,
\}\,
\label{rate2}
\end{equation}
where the coupling to a heat bath is reflected in the spectral function (\ref{sf}) whereas
the influence of the stochastic field on the $a\rightarrow b$ transition is concentrated  in the specific  \emph{stochastic field generated} (SFG) Lorentzian
%
\begin{equation}
L(\gamma,\nu;\Omega)=\frac{1}{\pi}\,
\frac{\gamma}{(\gamma -\Omega^2/\nu)^2
+\Omega^2}
\label{lor}
\end{equation}

It is important to stress that the rate constant (\ref{rate2}) includes a stochastic field (via the parameters $\nu$ and $\gamma\equiv\sigma^2/\nu$) in a non-perturbation manner. Therefore,  it becomes possible to analyze various stochastic regimes of the transition processes in macromolecules  dependending on the  value of the generic field parameters $\nu$ and $\gamma$.

Before applying the  theory to describe the  conformational quasi-isoenergetic transitions,  let us analyze  a situation when $|\Omega_{ab}\pm\omega_{\lambda}|\ll\nu$. Under this  condition,
the SFG Lorentzian $L(\gamma,\nu;\Omega)$ is reduced to a standard Lorentzian
%
\begin{equation}
L(\gamma,\Omega)=\frac{1}{\pi}\,
\frac{\gamma}{\gamma^2
+\Omega^2}\,
\label{lorr}
\end{equation}
with $\gamma=\gamma_D$, cf. Eq. (\ref{dem1}). It is necessary to
note that form (\ref{lorr}) is also valid in the case of TP,  if
only the function $F(\tau)$ demonstrates a single exponential drop
with decay parameter $\gamma=\gamma_T$, cf. Eq. (\ref{dem2}). In
Lorenzian  (\ref{lorr}), quantity $2\hbar\gamma$ can be treated as
the broadening of the molecular energy levels. If the  parameter
$\gamma$ is so small that   Lorentzian  (\ref{lorr}) appears as a
narrow peak,  one can calculate the rate constant taking a limit
$\gamma\rightarrow 0$. This converts the  Lorentzian into the
Dirac's delta-function $\delta (\Omega)$ thus reducing the rate
constant (\ref{rate2}) to
%
\begin{equation}
K_{ab}=|V_{ab}|^2\,\frac{J_{ab}(|\Omega_{ab}|)}{\hbar^4\Omega_{ab}^2}\,
\{[n(\Omega_{ab})+1]\theta(\Omega_{ab}) + n(\Omega_{ba})\theta(\Omega_{ba})\}\,.
\label{rate3a}
\end{equation}
[In Eq. (\ref{rate3a}), $\theta(\Omega)$ is the Heaviside unit function.] Note that in the case of DP, the form (\ref{rate3a}) is realized only in the limit of high frequency stochastic field (when both inequalities, $\nu_D\gg\sigma_D$ and $\nu_D\gg |\Omega_{ab}\pm\omega_{\lambda}|$, are satisfied for each bath mode  $\lambda$). In the case of TP, the same expression works under the condition $(\nu_T+\sigma_T^2/\nu_T)\gg |\Omega_{ab}\pm\omega_{\lambda}|$, i.e. in the two opposite limits of either  high frequency stochastic field ($\nu_T\gg \sigma_T$, the adiabatic regime) or low frequency stochastic field ($\nu_T\ll \sigma_T$, the quasi-static regime). For exoergic transitions, when
$\Omega_{ab}>0$ and $\exp{(\hbar\Omega_{ab}/k_BT)}\gg 1$, the rate constant (\ref{rate3a}) is independent of temperature attaining the optical limit $K_{ab}^{(opt)}=(|V_{ab}|^2/\hbar^4\Omega_{ab}^2)\,
J_{ab}(\Omega_{ab})$. Conversely, in the endoergic case when $\Omega_{ab}<0$ and $\exp{(-\hbar|\Omega_{ab}|/k_BT)}\ll 1$, the rate constant appears as
$K_{ab}^{(act)}=K_{ab}^{(opt)}\,\exp{(-E_{ab}^{(act)}/k_BT)}$, thus demonstrating the Arrhenius like law with $E_{ab}^{(act)}=\hbar |\Omega_{ab}|$ being the activation energy.

Now let us consider the  formation of the rate constant at the quasi-isoenergetic transitions ($\Omega_{ab}\approx 0$). To this end, one has to analyze the action of the moderately high frequency stochastic field. For such field, the condition $\gamma\gg |\Omega_{ab}\pm\omega_{\lambda}|$ is satisfied, and thus, the  SFG Lorentzian (\ref{lor}) is transformed into
%
\begin{equation}
L(\gamma,\Omega\approx 0)=\frac{1}{\pi\gamma}\,.
\label{lorr1}
\end{equation}
Note a fundamental fact that at the quasi-isoenergetic transitions, forms (\ref{lorr}) and (\ref{lorr1}) are valid not only for the  condition $\nu_D^2\gg \Omega^2$ but for the condition $\sigma_D^2\gg\Omega^2$ as well. This means that during the DP, the expression (\ref{lorr1}) can be used to describe the  quasi-isoenergetic transitions at any relation between the  amplitude and the frequency of the moderately high frequency stochastic field. Similar statement is true for the TP.

The room temperatures correspond to the energies of the order 0.025 eV, identical to $6\cdot 10^{12}$s$^{-1}$ or 200 cm$^{-1}$. Therefore, if the transitions are accompanied by the vibrations of the order 60 cm$^{-1}$ and lower, one can set $n(\omega_{\lambda})\simeq k_BT/\hbar\omega_{\lambda}$. This reduces the averaged rate constant (\ref{rate2}) to the form
%
\begin{equation}
K_{ab}=\frac{2|V_{ab}|^2\,k_BT}{\pi\,\hbar^2\gamma}\,\int_0^{\infty}\,d\omega
\frac{J_{ab}(\omega)}{\hbar^3\omega^3}\,.
\label{rate3}
\end{equation}

The physical origin of the stochastic parameter $\gamma$ is dictated by the random shifts of the  molecular energy levels.  We propose a phenomenological model where for the DP, the mean positive and the mean negative amplitudes ($+\sigma_D/2$  and $-\sigma_D/2$, respectively) are determined by  the relation    $[(2\pi\hbar\sigma_D)/2]^2=\overline{\delta E^2}$ with $\overline{\delta E^2}$ being the average of the square of energy fluctuations. In line with the general theory of thermodynamic fluctuations in canonic ensembles,  this average is calculated as $ \overline{\delta E^2}=k_BT^2(\partial \overline{E}/\partial T)$ \cite{ hil56} where $\overline{E}$ is the mean energy.
It is possible to show that in thermodynamic equilibrium, the average linear frequency of the  fluctuations is associated  with the average energy $\overline{E}$ so that $\nu_D=\overline{E}/2\pi\hbar$ \cite{fer06}. Thus, a stochastic parameter can be estimated with the  relation $\gamma_D=(2k_BT^2/\pi\hbar)(\partial \ln{\overline{E}}/\partial T)$. In the classical limit under consideration, the mean energy per a separate degree of freedom is equal \emph{exactly}  to thermal energy $k_BT$. Therefore, independently of the  precise molecular system, the mean  energy $\overline{E}$ is proportional to  the  bath temperature. It yields $\gamma_D=2k_BT/\pi\hbar$. If TP is formed as the sum of the two DP, then $\nu_T=2\nu_D$, $\sigma_T=\sigma_D$ and thus $\gamma_T=(4/3)[\sigma_D^2\nu_D/(4\nu_D^2 + \sigma_D^2)]=k_BT/3\pi\hbar$. In summary, the fast equilibrium  thermodynamic fluctuations form the stochastic decay parameter $\gamma=\tau_{stoch}^{-1}$ that obeys a linear dependence on temperature (in the classical limit). Thus,  one can generally set
$\gamma=\zeta\, (2k_BT/\pi\hbar)$ where factor $\zeta$ is determined by the type of stochastic process. [In the above examples, one derives $\zeta=1$ and $\zeta =1/6$ for the DP and the TP, respectively.] Consequently,
%
\begin{equation}
K_{ab}=\frac{2\pi}{\hbar}\,|V_{ab}|^2\,\Lambda_{ab}\,.
\label{rate4}
\end{equation}
In Eq. (\ref{rate4}), $\Lambda_{ab}=(2\pi\zeta)^{-1}\int_{0}^{\infty}\,
d\omega\,(J_{ab}(\omega)/\hbar^3\omega^3)$ is the temperature-independent factor.

To demonstrate how the temperature dependence of the averaged rate constant (\ref{rate2}) is canceled, we specify the form of a spectral function using the Song-Marcus's model \cite{son93,tan99}. For such model,  $J_{ab}(\omega)=2\pi\hbar\omega\,E^{(reorg)}_{ab}\,\delta (\omega-\omega_0)$ where $\omega_0$ is the characteristic frequency of the bath phonon and
$E^{(reorg)}_{ab}=(g^{(0)}_{ab})^2\,\hbar\omega_{0}$ is the reorganization energy of the $a\rightarrow b$ transition with participation of the only vibrational mode. The substitution
$J_{ab}(\omega)$ in Eq. (\ref{rate2}) reduces rate constant to the following reading form
%
\begin{equation}
K_{ab}=K_{ab}^{(0)}\,(\eta_- +\eta_+)\,.
\label{ratemar}
\end{equation}
Here, $K_{ab}^{(0)}\equiv(2\pi|V_{ab}|^2/\hbar^3\omega_0^2)\,E^{(reorg)}_{ab}$ is the temperature-independent part of the rate whereas
%
\begin{equation}
\eta_-=\omega_0\,(1+n(\omega_0))\,L(\gamma,\nu;\Omega_{ab}-\omega_{0})
\label{temfact-}
\end{equation}
and
%
\begin{equation}
\eta_+=\omega_0\,n(\omega_0)\,L(\gamma,\nu;\Omega_{ab}+\omega_0)\,
\label{temfact+}
\end{equation}
are the temperature-determining factors.

\begin{figure}
\includegraphics[width=8cm]{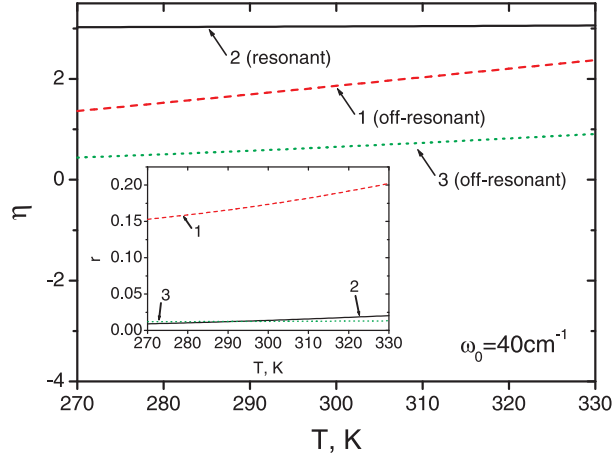}
\caption{Temperature behavior of the averaged transition rate constant (\protect\ref{ratemar}) at dichotomous action of stochastic field. The physiologically important temperature region is presented. Curves 1 and 3 correspond to the off-resonant transition frequencies $\Omega_{ab}= 10$ cm$^{-1}$ and $\Omega_{ab}= 80 $cm$^{-1}$, respectively. Curve 2 manifests the cancellation of the  temperature dependence at a resonant regime of transition at which transition frequency
$\Omega_{ab}$= 40 cm$^{-1}$ coincides with  phonon frequency $\omega_{0}$.  The insertion shows the ratio $r=\eta_+/\eta_-$ of the ``off-resonant'' and the ``resonant'' terms in the  transition rate. The estimation of the factor $\eta\equiv K_{ab}/K_{ab}^{(0)}=\eta_- + \eta_{+}$ is  based on Eqs. (\protect\ref{lor}), \protect(\ref{temfact-})  and (\protect\ref{temfact+}) at $\gamma =2k_BT/\pi\hbar$ and  $\nu=k_BT/2\pi\hbar$.
}
\label{fig1}
\end{figure}

In Fig. \ref{fig1}, the temperature dependence of  the  transition rate constant is presented.
When the transition frequency $\Omega_{ab}$ is far from the phonon frequency $\omega_0$ (``off-resonant'' regime of transition) the factors $\eta_-$ and $\eta_+$ are proportional to temperature (1 and 3 curves). The situation is changed significantly at the quasi-isoenergetic (``resonant'') regime of transition when $\Omega_{ab}\simeq\omega_0$. In this case, the main contribution to the rate constant is determined by the factor $\eta_-$ which now noticably exceeds the nonresonant factor $\eta_+$ (cf. insertion to Fig. \ref{fig1}). But, at the quasi-isoenergetic transition, due to the  linear temperature rise of the stochastic  parameter $\gamma$ the  factor $\eta_-$ demonstrates a practically temperature-independent behavior (compare curve 2 with  curves 1 and 3). In fact, one can talk  about a temperature independent character of the transition process.

\section{Experimental support}

Eq. (\ref{rate4}) and  Fig. \ref{fig1} present an important result, namely, that fast thermodynamic fluctuations are able to cancel the temperature dependence of rate constants characterizing the quasi-isoenergetic transitions in the  macromolecular systems. In this section, we discuss the supporting experimental findings related to  transfer processes in biological structures.
In the begining of the Section 2, we have already noted the results concerning the observation of energy fluctuations in biomacromolecules. Here, we note an important finding that
the fluorescence of the flavin cofactor indicates  ``all-or-none'' enzymatic turnovers of the flavin reductase \cite{lu98,yan03} while energy fluctuations in the pigment-protein comlexes  of the  light-harvesting antenna from photosynthetic bacteria  exhibit the dichotomous behavior  \cite{val07,jan08}. It is quite possible to extrapolate these findings to  other protein containing complexes. Thus, the kinetic equations (\ref{me4}) and respective  rate constants (\ref{rate2}) derived by taking into account the averaging over the discrete energy fluctuations  in a macromolecule, can be applied to explain the following  experimental results.

\begin{figure}
\includegraphics[width=8cm]{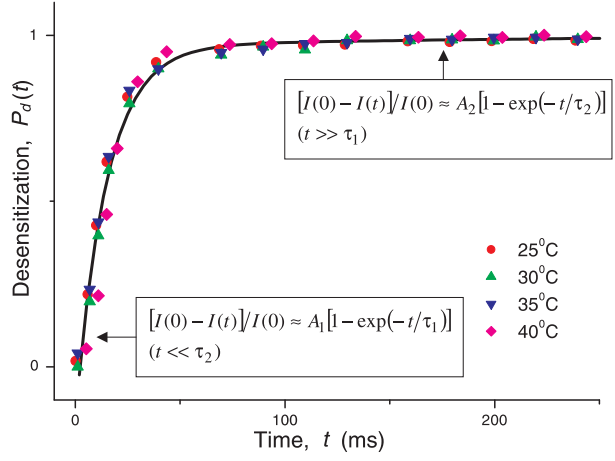}
\caption{Temperature independence of the desensitization  onset of P2X$_3$ receptors measured at physiological temperatures 25-40$^{\circ}$C (selected from \protect\cite{khm08}) and its fit by Eq. (\ref{exp}).} \label{fig2}
\end{figure}

\emph{Example 1-Temperature independence of desensitization onset of
P2X$_3$ receptors.}
The P2X$_3$ receptors belong to the family of  ionotropic receptors widely evolved in the peripheral nervous system. These receptors are highly specific membrane proteins which link the binding of ATP molecules and their analogues to the opening and closing of the gate of a selective transmembrane ion pore \cite{kha01}. The experimental data  for the P2X$_3$ receptors were obtained at phisiologically important temperatures of 25, 30, 35, and 40$^{\circ}$C \cite{khm08}. [In Fig \ref{fig2}, these data  are reproduced with different symbols.] We have shown that  the ATP-activated currents $I(t)/I(0)$ measured at different temperatures, manifest an identical  two-stage decrease. To our knowledge, this is the first quantitative observation of a temperature-independent gating in biological membranes. Evolution of the current is well described by the  theoretical curve reflecting the  double-exponential transition kinetics,
%
\begin{equation}
P_d(t)=1-A_1\,e^{-t/\tau_1}-A_2\,e^{-t/\tau_2}\,,
\label{exp}
\end{equation}
with $P_d(t)=1-I(t)/I(0)$ being the desensitization probability of the channels. In Eq. (\ref{exp}), $A_1\simeq 0.968$ and $A_2\simeq 0.032$ are the pre-exponential weights whereas $\tau_1\simeq 14.7$ms and $\tau_2\simeq 231$ms are the temperature-independent characteristic times of the desensitization.

Despite that  primary, secondary and tertiary molecular structure of
P2X$_3$ receptors is well-resolved \cite{kha01}, the particular molecular content of the  receptor's gate(s) is yet unknown. But, it is possible to deduct that because of the hydrogen and non-covalent (ionic) bond vibrations occur with characteristic frequencies 0.2-20 THz \cite{bak00,tad04}, the fast local fluctuations of channel structures can not directly lead to the slow open-closed transitions accompanied by extended flexible changes.  However, these local fluctuations can create the stochastic fields varying the position of energy levels.

We interpret the above experimental result in the framework of a model where the formation of an ion current appears as the sum of two partial currents.
Let  $i_j$ be the current through a separate open channel of the  $j(=1,2)$th type. If   $N_j$ is the number  of respective ion channels, then a partial current  reads as $I_j(t)=N_j\,i_j\,P^{(j)}_{o}(t)$ where $P^{(j)}_{o}(t)$ is the probability for the channel to be in  an open state "$o$". Desensitization of each channel occurs independently  reflecting the process of conformational transition from the open (conductive) state into the closed (nonconductive) state so that  $P^{(j)}_{o}(t)=\exp{(-t/\tau_j)}$. Introducing the quantities  $I(0)\equiv \sum_{j=1,2}\,N_j\,i_j\,$  and  $A_1=\xi/(1+\xi)$ and $A_2=1/(1+\xi)$ where
$\xi\equiv N_1\,i_1\,/N_2\,i_2\,$, one  expresses an ion current in the form $I(t)=I(0)\,P_o(t)$ where $P_o(t)=\sum_{j=1,2}\,A_j\,P^{(j)}_{o}(t)$ is the apparent (statistically averaged) probability of an ion channel to be in an open state. Bearing in mind that $P_{d}(t)=1-P_{o}(t)$, one arrives at the Eq. (\ref{exp}) (see also \cite{foot}).
The physical explanation of the temperature independency of the desensitization process can be given based on the above proposed model of quasi-isoenergetic transitions in flexible molecules.
Denoting via $a=jo\alpha$ and
$b=jd\beta$ the conformational isoenergetic  substates participating in the open-close  transition, we introduce the integral occupancies of states $o$ and $d$ as $P_o^{(j)}(t)=\sum_{\alpha=1}^{\mu_{jo}}\,P_{jo\alpha}(t)$ and $P_d^{(j)}(t)=\sum_{\beta=1}^{\mu_{jd}}\,P_{jd\beta}(t)=1-P_o^{(j)}(t)$ where $\mu_{jo}$ and $\mu_{jd}$ are the numbers of the respective degenerated substates for the $j$th type of channel.
Eq. (\ref{me4}) now reads
%
\begin{equation}
\dot{P}_o^{(j)}(t)=-\big[K^{(j)}_{o\rightarrow d}P_o^{(j)}(t)-K^{(j)}_{d\rightarrow o}P_d^{(j)}(t)\big]
\label{me5}
\end{equation}
where
%
\begin{equation}
 K^{(j)}_{o\rightarrow d}=\frac{1}{\mu_{jo}}\,K_j \,
\label{rate5}
\end{equation}
is the forward rate constant and (cf. Eq. (\ref{rate4}))
%
\begin{equation}
K_j\equiv \frac{2\pi}{\hbar}\,|V_{od}|^2\,\sum_{\alpha=1}^{\mu_{jo}}\,
\sum_{\beta=1}^{\mu_{jd}}\,\Lambda_{jo\alpha\,jd\beta}\,.
\label{rate5a}
\end{equation}
[The backward rate constant is  given by expression $K^{(j)}_{d\rightarrow o}=K_j/\mu_{jd} $.]

Since the disorder characteristic for desensitization increases the degeneracy of the molecular state, then
$\mu_{jd}\gg \mu_{jo}$. This reduces the kinetics of quasi-isoenergetic transitions to the nonrecurrent single-exponential kinetics
with the characteristic time $\tau_j\simeq (K^{(j)}_{o\rightarrow d})^{-1}$ for the $j$th type of channel. Thus, physically, the receptor desensitization appears as a temperature-independent open$\rightarrow$close transition process between the degenerated quasi-isoenergetic molecular conformations of an ion channel.

\emph{Example 2.-Temperature-independent degradation of an endogenous PER2.
} Very recently, a direct observation of  temperature-independent conformational transformations in proteins has been provided in living clock cells \cite{iso09}. It is known that a circadian periodicity in different organisms ranging from bacteria to mammals remains robust over a wide range of temperatures (see e.g. \cite{rep02}). One explanation of this characteristic may be that it is due to the temperature insensitivity of the degradation of PER2 protein which is the key period-determining protein in the mammalian circadian clock cascade. It has been shown that the phosphorylated PER2 degrades with a particular rate which was extremely sensitive to chemical perturbation, however, was remarkably unaffected by physical perturbation such as temperature shifts \cite{iso09}. \textit{In vivo}, the degradation of an endogenous PER2 is regulated by casein kinase I$_\epsilon$-dependent phosphorylation. During the  phosphorylation, the ATP molecule is hydrolized thus donating inorganic phosphate oxygen groups to the chemical structure of PER2. These groups, being negatively charged, can generate high frequency hydrogen bond fluctuations which, in turn, will affect the protein degradation related to slow conformation motions. Therefore, as in the previous case of P2X$_3$ receptors, the model of the stochastic regulation of the transition rate is also applicable.

\begin{figure}
\includegraphics[width=8cm]{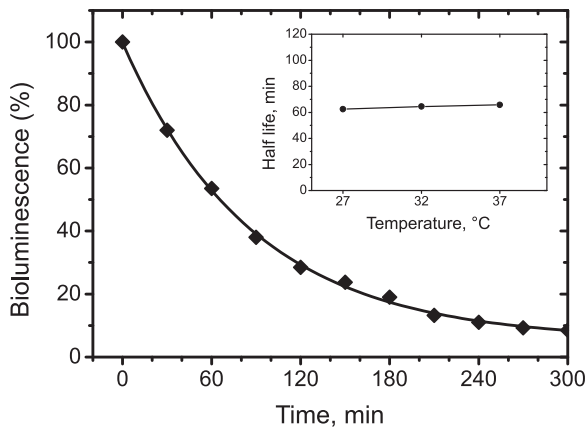}
\caption{The temperature independence of the degradation of PER2 protein in mouse embrionic fibroblasts (reproduced from Figs. 2D and 2E of ref. \protect\cite{iso09}). The degradation is monitored as a decay of luminescence (in percentages) of the PER2-luciferase complex. The diamonds indicate the data at 37$^{\circ}$C and the line represents an exponential curve. The insertion depicts the averaged life times  calculated from the approximated curves at 27, 32 and 37$^{\circ}$C.}
\label{fig3}
\end{figure}

Experimentally, the PER2 degradation can be monitored by a decay of the bioluminescence (cf. Fig. \ref{fig3})
and appears as a single exponential drop of the amplitude of bioluminescence signal, $A(t)=\exp{(-k_{deg}t)}$ where $k_{deg}\approx 1.7\cdot 10^{-3}$s$^{-1}$ is the slow degradation rate constant. As one can see, the half-times of the decay curves has negligible temperature dependence (see the insertion to Fig.\ref{fig3}) and may be considered temperature-insensitive. The physical mechanism of such temperature independency in physiologically important temperature region, can be also associated with the equilibrium thermodynamic fluctuations and a pronounced disorder in the protein structure.  This reduces the kinetics of the quasi-isoenergetic transitions to the nonrecurrent kinetics with the temperature-independent rate constant $k_{deg}$  similar to that is given by Eq. (\ref{rate5}).

\section{Conclusions}

In this communication, a stochastic master equation (\ref{deq}) for the
diagonal part of the density matrix $\rho_d(t)$ of an  open quantum
system is derived under the assumption  that state energies $E_a(t)$
are affected by random time-dependent fields. Special attention is given to derivation of kinetic equations for the averaged state occupancies of a macromolecule, Eqs. (\ref{me2}) and (\ref{me4}), and respective averaged rate constants, Eq.  (\ref{rate1}). The derivation is based on the model that incorporates two different types of motion in macromolecular system.
The first type of motion is associated with a heat bath. The energy of the phonons belonging to a heat bath, covers the energy difference between the molecular states involved in the transition. Therefore, the bath phonons directly participate in the transition. The second one is characterized by the high frequency vibrations. The energy of the respective phonons strongly exceeds  the transition energy, so that the high frequency vibrations do not directly participate in the  transition. At the same time, the high frequency vibrations can create stochastic fields varying the energy of molecular states. The details of the \emph{exact} averaged procedure are demonstrated for two types of Kubo-Anderson  stochastic processes, the dichotomous and the trichotomous ones. We found the
characteristic times $\tau_{stoch}^{(j)}$ of the drop of the averaged
stochastic functional $F(\tau)$. Additionally,  the conditions when
the $F(\tau)$ exhibits a single-exponential drop, have been  established.
Special attention was  focused  on the analysis of the quasi-isoenergetic transitions. Here,
a novel physical mechanism was proposed to explain the formation of the temperature-independent transition processes in molecular systems. In contrast with the well established mechanism describing a quantum phononless site-to-site low temperature particle tunneling or a tunneling at room temperatures that is accompanied by an emission of  high frequency phonons (when frequency $\omega_{\lambda}$ of the separate phonon satisfies the condition  $\omega_{\lambda}\gg  k_BT/\hbar$) \cite{jor76,jor80}), the mechanism proposed here works in a classic (physiological) region of temperatures and at a weak coupling to the bath phonons. At such conditions, the  transition  is accompanied by creation or annihilation of a single low frequency phonon of the frequency $\omega_{\lambda}\ll k_BT/\hbar$.   The lack of temperature dependence in  rate constant (\ref{rate2}) occurs due to the thermodynamical stochastic variation of energy levels participating in the transitions. The frequency of this variation has to be of the order of the characteristic thermal frequency $k_BT/\hbar$. This reduces
an averaged rate constant (\ref{rate2}) to a  more simple form (\ref{rate3}) which, in turn, is simplified to the temperature-independent form (\ref{rate4}). The two recent experimental observations  of temperature-independent transitions in biosystems, are discussed. We show that these results can be explained in the framework of the model where the transitions are performed between
the quasi-isoenergetic conformation states of macromolecules in the presence of random thermodynamic fluctuations of respective energy levels. In this case, both degradation and desensitization of proteins are generally accompanied  by a pronounced disorder in the protein structure thus essentially increasing the degeneracy of the quasi-isoenergetic molecular
states. The explanation of the experimental results is based  on the model where part of molecular motions refer to fluctuations. Note that the presence of the stochastic motions in biological macromolecules is supported by spectroscopic measurements, thus indicating the existence of energy fluctuations  of all-or-none type in protein containing complexes \cite{lu98,yan03,val07,jan08}.

\section{Acknowledgments} This paper is dedicated to the P. H\"anggi jubilee. One of the authors E.P. thanks Peter H\"anggi for the fruitful long-term scientific collaboration shared contributions to quantum kinetics and molecular electronics.

                                                                           %


\begin{thebibliography}{00}
\bibitem{foota} The form of interaction $V$ is dictated by a concrete transition process. In our case, operator $V$ refers to the ``coated''  off-diagonal interaction responsible for the transitions between the molecular states, cf. Eq. (\protect\ref{trop}).
\bibitem{foot} Analogous form arises in the one-type channel model  with two open
channel configurations of different conductivity. In this case, however, the weights $A_1$ and $A_2$ are formed before the desensitization starts and these weights remain unchanged during the desensitization. I.e. one has additionally to suppose that the transitions between the two noted channel configurations occur on the time scale which is much larger than the characteristic desensitization times $\tau_1$ and $\tau_2$.
\end{thebibliography}
\end{document}